\documentclass[prl,aps,twocolumn,showpacs]{revtex4}

\usepackage{amssymb}
\usepackage{mathrsfs}
\usepackage{amsmath}
\usepackage{graphicx}

\begin{document}
\title{Observation of Anti-correlation of Classical Chaotic Light}

\author{Hui Chen, Sanjit Karmakar, Zhenda Xie, and Yanhua
Shih}
\affiliation{Department of Physics, University of Maryland,
Baltimore County, Baltimore, MD 21250}

\begin{abstract}
We wish to report an experimental observation of anti-correlation from first-order
incoherent classical chaotic light.  We explain why the classical statistical theory
does not apply and provide a quantum interpretation.  In quantum theory,
either correlation or anti-correlation is a two-photon interference phenomenon, which
involves the superposition of two-photon amplitudes, a nonclassical entity corresponding
to different yet indistinguishable alternative ways of producing a joint-photodetection
event.
\end{abstract}
\pacs{PACS Number: 03.65.Bz, 42.50.Dv}
\maketitle

\begin{figure}[hbt]
\centering
\includegraphics[width=60mm]{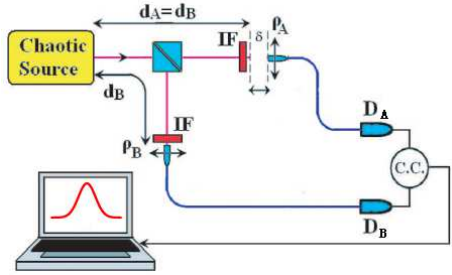}
\caption{Schematic of a modern HBT interferometer which measures
both temporal and spatial correlation of light by scanning the
optical fiber tips longitudinally or transversely.} \label{fig:HBT}
\end{figure}

In 1956, Hanbury Brown and Twiss (HBT) discovered a nontrivial
intensity correlation in thermal light \cite{HBT}.
Figure~\ref{fig:HBT} schematically illustrates a modern HBT
interferometer or the so called ``intensity interferometer".   In a
temporal HBT interferometer, the temporal, randomly distributed,
chaotic thermal light has a twice greater chance of being measured
within its coherence time by the joint-detection of two individual photodetectors.
In a spatial HBT interferometer, the spatial, randomly distributed, chaotic
thermal light exhibits a twice greater chance of being captured within
a small transverse area that equals the spatial coherence of the thermal
radiation by two point photodetectors . It was recently found that for a
large angular sized chaotic source the spatial correlation is effectively
within a physical ``point".  The point-to-point near-field spatial correlation
of chaotic light has been utilized for reproducing nonlocal ghost
images in a lensless configuration \cite{Gpaper}.

What is the physical cause of this peculiar behavior of chaotic thermal
light? In the classical point of view, the HBT phenomenon is a
statistical correlation of intensity fluctuations of thermal radiation.
In general, no matter how complicated the optical setup is,
classical theory considers the joint-detection between two individual,
point-like photodetectors, $D_A$ and $D_B$, as a
measure of the statistical correlation of two intensities at space-time
coordinates $(\mathbf{r}_A, t_A)$ and $(\mathbf{r}_B, t_B)$
\begin{align}\label{Classical-0}
& \ \ \ \ \Gamma^{(2)}(\mathbf{r}_A, t_A; \mathbf{r}_B, t_B)
= \langle I(\mathbf{r}_A, t_A) \, I(\mathbf{r}_B, t_B) \rangle \\ \nonumber
&= \langle I(\mathbf{r}_A, t_A) \rangle\langle I(\mathbf{r}_B, t_B) \rangle
+ \langle \Delta  I(\mathbf{r}_A, t_A) \, \Delta I(\mathbf{r}_B, t_B) \rangle.
\end{align}
The HBT observation comes from the second term of
Eq.~(\ref{Classical-0}), which indicates correlated intensity
fluctuations at a distance. Why do the two distant intensities fluctuate
in such a peculiar manner?  One historical answer is
``photon bunching", i.e., a thermal light source has a higher chance
of emitting photons in pairs.  Another involves the coherence
of the electromagnetic fields
\begin{align}\label{Classical-1}
& \ \ \ \ \Gamma^{(2)}(\mathbf{r}_A, t_A; \mathbf{r}_B, t_B)
= \langle I(\mathbf{r}_A, t_A) \, I(\mathbf{r}_B, t_B) \rangle \nonumber \\
& = \langle E^{*}(\mathbf{r}_A, t_A)E(\mathbf{r}_A, t_A)
E^{*}(\mathbf{r}_B, t_B)E(\mathbf{r}_B, t_B) \rangle \nonumber \\
&= \Gamma^{(1)}_{AA}\Gamma^{(1)}_{BB} + \Gamma^{(1)}_{AB}\Gamma^{(1)}_{BA},
\end{align}
where $\Gamma^{(1)}_{jj} =  \langle E^{*}(\mathbf{r}_j, t_j)E(\mathbf{r}_j, t_j)
\rangle =  \langle I(\mathbf{r}_j, t_j) \rangle$, $j = A, B$,
is defined as the self-coherence function, or self-correlation function of the field;
$\Gamma^{(1)}_{AB} = \Gamma^{*(1)}_{BA} = \langle E^{*}(\mathbf{r}_A, t_A)E(\mathbf{r}_B, t_B) \rangle$
is defined as the mutual-coherence, or mutual-correlation function of the field.
It has been well accepted that an intensity-interferometer measures the
mutual-correlation $\Gamma^{(1)}_{AB}$ of the input field, no matter how
complicated the optical setup.

\begin{figure}[hbt]
\centering
\includegraphics[width=80mm]{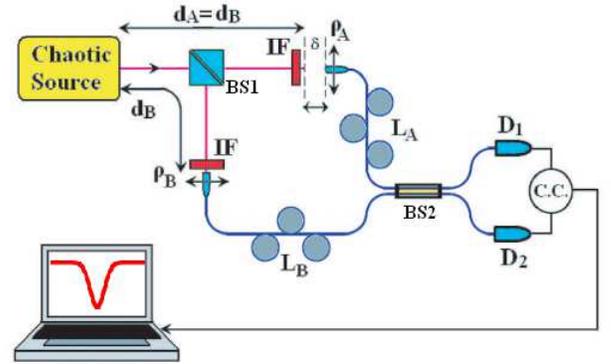}
\caption{Schematic setup of the experiment.} \label{fig:scheme}
\end{figure}
Now we consider a slightly different experimental setup in
Fig.~\ref{fig:scheme}. Instead of directly measuring the intensity
correlation of $\langle I(\mathbf{r}_A, t_A) \, I(\mathbf{r}_B, t_B)
\rangle$, this setup measures  $\langle I(\mathbf{r}_1, t_1) \,
I(\mathbf{r}_2, t_2) \rangle$ at the output ports of a $50/50$
optical fiber beamsplitter.  If the two input fiber tips  A and B
are placed within the longitudinal coherence time and the transverse
coherence area of the thermal field, this setup is equivalent to a
Mach-Zehnder interferometer.  $D_1$ and $D_2$ will each measure
first-order interference as a function of the optical delay $\delta$
when scanning the fiber tip A along its longitudinal axis.
Consequently, the joint-detection of $D_1$ and $D_2$ outputs an
interference pattern that is factorizable into two first-order
interferences. However, this is not our desired experimental
condition.   We decided to move the fiber tip A outside the
transverse coherence area to force $\Gamma^{(1)}_{AB} = 0$. Under
this condition, there would be no observable first-order interference in
a timely averaged measurement and in a instantaneous
``single-exposure" observation \cite{Mandel}. Consequently, the
standard classical statistical correlation theory of intensity fluctuations gives
the following second-order correlation $\Gamma^{(2)}$ that is independent of
optical delay $\delta$,
\begin{align}\label{Classical-00}
&\ \ \ \ \Gamma^{(2)}(\mathbf{r}_1, t_1; \mathbf{r}_2, t_2)
=\langle I(\mathbf{r}_1, t_1) \, I(\mathbf{r}_2, t_2) \rangle  \\ \nonumber
&=\langle \big{|} E(\mathbf{r}_A, t_{1A}) + E(\mathbf{r}_B, t_{1B})
\big{|}^2
\big{|} E(\mathbf{r}_A, t_{2A}) - E(\mathbf{r}_B, t_{2B}) \big{|}^2 \rangle,
\end{align}
where $(\mathbf{r}_A, t_{jA} = t_j - r_{Aj}/c)$ and $(\mathbf{r}_B,
t_{jB} = t_j - r_{Bj}/c$), $j = 1,2$, with $r_{Aj}/c$ ($r_{Bj}/c$)
the optical delay from point A (B) to $D_j$, are earlier space-time
coordinates at points $A$ and $B$. In the Appendix A we show in
detail why the classical statistical intensity correlation
$\Gamma^{(2)}(\mathbf{r}_1, t_1; \mathbf{r}_2, t_2)$ is $\delta$
independent.  In fact, the physics is rather simple; mathematically
Eq.~(\ref{Classical-00}) has sixteen terms to begin with.  However,
all $\Gamma^{*(1)}_{AB}\Gamma^{(1)}_{AB}$ terms vanish due to the
mutual incoherence between fields $E(\mathbf{r}_A, t_A)$ and
$E(\mathbf{r}_B, t_B)$, leaving only these nonzero terms which have
no contribution from the $\delta$-dependent first-order mutual
coherence.

\begin{figure}[hbt]
\centering
\includegraphics[width=80mm]{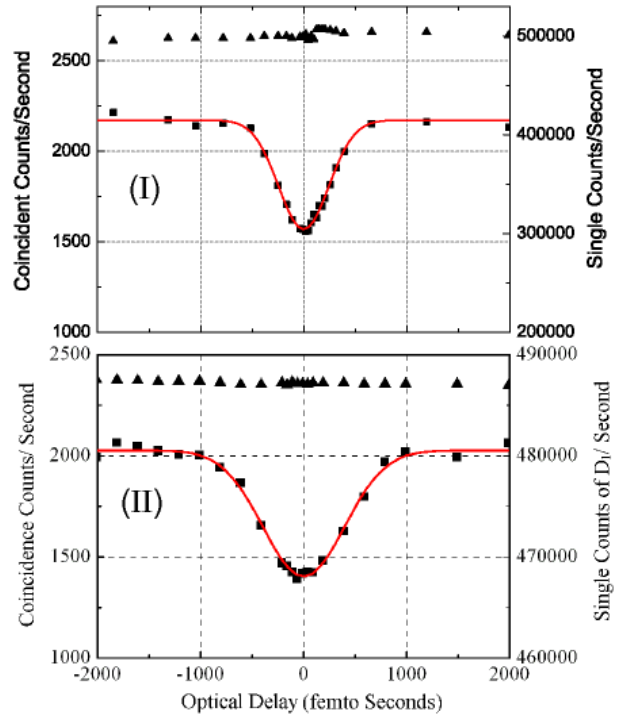}
\caption{Two typical observed anti-correlation functions with
different coherence time of the chaotic field.  $\tau_c \sim 345 fs$
for (I), $\tau_c \sim 541 fs$ for (II).   The coherence time is
determined by the bandwidth of the spectral filters (IF).
}\label{fig:anti-correlation}
\end{figure}

The measurement produces quite a surprise. A ``unexpected"
anti-correlation ``dip" was observed in the joint-detection counting
rate of $D_1$ and $D_2$ as a function of the optical delay $\delta$.
Figure~\ref{fig:anti-correlation} reports two typical measured
anti-correlation functions with different spectrum bandwidths of the
chaotic field.

The experimental detail is described as follows.

(1) The source: the light source is a standard pseudo-thermal source
that was developed in the 1960's and used widely in HBT correlation
measurements \cite{Pseudo}. The source consists of a CW mode-locked
laser beam with $\sim$200 femtosecond pulses at a 78~MHz repetition
rate and a fast rotating diffusing ground glass. The linearly
polarized laser beam is enlarged transversely onto the ground glass
with a diameter of $4.5 mm$. The enlarged laser radiation is
scattered and diffused by the rotating ground glass to simulate a
near-field, chaotic-thermal radiation source: a large number of
independent point sub-sources with independent, random, relative
phases.

(2) The interferometer: A 50/50 beam splitter (BS1) is used to split
the chaotic light into transmitted and reflected radiations which
are then coupled into two identical polarization-controlled
single-mode fibers A and B respectively. The fiber tips are located
$\sim200mm$ from the ground glass, i.e. $d_A=d_B\sim200mm$. At this
distance, the angular size $\Delta\theta$ is $\sim22.5$ mili-radian
($1.29^\circ$) with respect to each input fiber tip, which satisfies
the Fresnel near-field condition. Two identical narrow-band spectral
filters (IF) are placed in front of the two fiber tips A and B. The
transverse and longitudinal coordinates of the input fiber tips are
both scannable by step-motors.  The output ends of the two fibers
can be directly coupled into the photon counting detectors $D_1$ and
$D_2$, respectively, for near-field HBT correlation measurements, or
coupled into the two input ports of a 50/50 single-mode optical
fiber beamsplitter(BS2) for the anti-correlation measurement.

(3) The measurement:  two steps of measurements were made. The
purpose of step one is to confirm the light source produces
chaotic-thermal field.  We measured the HBT temporal and spatial
correlation by scanning the input fiber tips longitudinally and
transversely.  In this measurement the output ends of the fibers are
coupled into $D_A$ and $D_B$ directly as shown in
Fig.~\ref{fig:HBT}. Chaotic radiation can easily be distinguished
from a laser beam by examining its second-order coherence function
$G^{(2)}(\mathbf{r}_A, t_A; \mathbf{r}_B, t_B)$, which is
characterized experimentally by the coincidence counting rate that
counts the joint photo-detection events at space-time points
$(\mathbf{r}_A, t_A)$ and $(\mathbf{r}_B, t_B)$.
\begin{figure}[hbt]
\centering
\includegraphics[width=70mm]{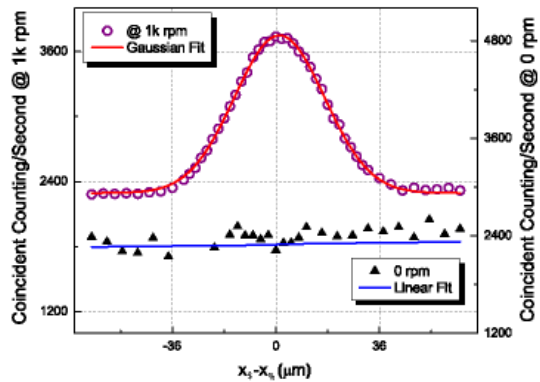}
\caption{Measurement of $G^{(2)}(x_A - x_B )$ at no rotation(0 rpm)
and at 1000 rpm.  Here, $x_A$ and $x_B$ are the x-component of
$\vec{\rho}_A$ and $\vec{\rho}_B$, and correspondingly y components
are kept $y_A=y_B$.} \label{fig:Speed}
\end{figure}
Figure~\ref{fig:Speed} reports two measured second-order
correlations at zero and at 1000 revolutions per minute (rpm) of the
rotating ground glass. This measurement guarantees a typical HBT
correlation of chaotic light at rotation speeds greater than 1000
rpm of the ground glass, indicating the chaotic nature of the light
source.  In this measurement, we have also located experimentally
the longitudinal and transverse coordinates of the fiber tips A and
B for achieving the maximum coincidence counting rate, corresponding
to the maximum second-order correlation.  In step two, we couple the
50/50 fiber beamsplitter (BS2) into the setup as shown in
Fig.~\ref{fig:scheme}.  This measurement was done in two steps.  We
first measured the first-order interference at $\vec{\rho}_A =
\vec{\rho}_B$ by scanning the input fiber tip $A$ longitudinally in
the neighborhood of $d_A \sim200$mm. There is no surprise to have
first-order interference in the counting rates of $D_1$ and $D_2$
respectively.  When choosing $\vec{\rho}_A = \vec{\rho}_B$, the two
input fiber tips are coupled within the spatial coherence area of
the radiation field; we have effectively built a Mach-Zehnder
interferometer.  We then move the input fiber tip $A$ transversely
from $\vec{\rho}_A = \vec{\rho}_B$ to $| \vec{\rho}_A - \vec{\rho}_B
| \gg l_c$, where $l_c$ is the transverse coherence length of the
thermal field. Then we scan the input fiber tip $A$ again
longitudinally in the neighborhood of $d_A \sim200$mm. The optical
delay between the plane $z = d_A \sim200$mm and the scanning input
fiber tip $A$ is labeled as $\delta$ in Fig.~\ref{fig:scheme}. We
have thus achieved the expected experimental condition of
$\Gamma^{(1)}_{AB} = 0$. There is no surprise we lose any
first-order interference in this experimental condition.  However,
it is indeed a surprise that in the joint-detection of $D_1$ and
$D_2$ an anti-correlation is observed as a function of the optical
delay $\delta$ that is reported in Fig.~\ref{fig:anti-correlation}.
In these measurements, $| \vec{\rho}_A - \vec{\rho}_B | \sim 40
l_c$.

Quantum theory gives a reasonable explanation to this surprising
observation.   In quantum theory, the second-order
correlation function represents the probability of observing a
joint-photodetection event at space-time coordinates
$(\mathbf{r}_1, t_1)$ and $(\mathbf{r}_2, t_2)$ \cite{Glauber}.
If more than one, different yet indistinguishable, alternative ways
of triggering a joint-photodetection event exist, these probability
amplitudes must be linearly superposed resulting in an interference.
The HBT correlation of chaotic light is the result of two-photon
interference, which involves the superposition of two-photon
amplitudes,
\begin{align}\label{Quantum-0}
& \ \ \ G^{(2)}(\mathbf{r}_1, t_1; \mathbf{r}_2, t_2) \\ \nonumber
&= tr{\big{[}\, \hat{\rho} \, \hat{E}^{(-)}(\mathbf{r}_1, t_1)
\hat{E}^{(-)}(\mathbf{r}_2, t_2) \hat{E}^{(+)} (\mathbf{r}_2, t_2)
\hat{E}^{(+)}(\mathbf{r}_1, t_1) \, \big{]}} \\ \nonumber &=
\sum_{\alpha,\beta} P_{\alpha\beta} \big{|}
\frac{1}{\sqrt{2}}\big{[} \mathscr{A}(\mathbf{r}_{\alpha 1}, t_1;
\mathbf{r}_{\beta 2}, t_2) + \mathscr{A}(\mathbf{r}_{\beta 1}, t_1;
\mathbf{r}_{\alpha 2}, t_2) \big{]} \big{|}^2,
\end{align}
where $\hat{E}^{(\pm)}(\mathbf{r}_j, t_j)$, $j = 1,2$, is the
positive ($+$) or negative ($-$) field operator at coordinate
($\mathbf{r}_j, t_j$). In Eq.~(\ref{Quantum-0}), we have treated the
chaotic radiation in a mixed state
\begin{align*}
\hat{\rho} \simeq |0 \rangle\langle 0 | + \sum_{\alpha} P_{\alpha} |
\Psi_{\alpha} \rangle \langle \Psi_{\alpha} | + \sum_{\alpha,\beta}
P_{\alpha\beta} | \Psi_{\alpha} \rangle |\Psi_{\beta} \rangle
\langle \Psi_{\beta} |\langle \Psi_{\alpha} | + ...,
\end{align*}
which represents an ensemble of sub-radiations, such as trillions of
photons created from a large number of independent and randomly
radiated sub-sources.  In the third term of $\hat{\rho}$, which
contributes to the joint-photodetection, $P_{\alpha\beta}$
represents the probability for the $\alpha$th and the $\beta$th
sub-radiations to be in the states $| \Psi_{\alpha}
\rangle|\Psi_{\beta} \rangle$. We have also defined an effective
wavefunction $G^{(2)}_{\alpha\beta} = |\Psi_{\alpha\beta}|^2$ with
\begin{equation}\label{symme}
\Psi_{\alpha\beta} = \frac{1}{\sqrt{2}}
\big{[}\mathscr{A}(\mathbf{r}_{\alpha 1}, t_1; \mathbf{r}_{\beta 2},
t_2) + \mathscr{A}(\mathbf{r}_{\beta 1}, t_1; \mathbf{r}_{\alpha 2},
t_2) \big{]}
\end{equation}
where $\mathscr{A}(\mathbf{r}_{\alpha 1}, t_1; \mathbf{r}_{\beta 2},
t_2) = \langle 0|\hat{E}_1^{(+)} |\Psi_\alpha \rangle \langle
0|\hat{E}_2^{(+)} |\Psi_\beta \rangle$ [or
$\mathscr{A}(\mathbf{r}_{\beta 1}, t_1; \mathbf{r}_{\alpha 2}, t_2)
= \langle 0|\hat{E}_1^{(+)} |\Psi_\beta \rangle \langle
0|\hat{E}_2^{(+)} |\Psi_\alpha \rangle$] represents the alternative
in which the $\alpha$th [or $\beta$th] photon and the $\beta$th [or
$\alpha$th] photon are annihilated at $(\mathbf{r}_1, t_1)$ and
$(\mathbf{r}_2, t_2)$, respectively. The symmitrized effective
wavefunction in Eq.~(\ref{symme}) plays the same role as that of the
symmitrized wavefunction of identical particles \cite{Bari}.
Obviously, it is the superposition of
$\mathscr{A}(\mathbf{r}_{\alpha 1}, t_1; \mathbf{r}_{\beta 2}, t_2)$
and $\mathscr{A}(\mathbf{r}_{\beta 1}, t_1; \mathbf{r}_{\alpha 2},
t_2)$ causing the nontrivial HBT correlation.

In the view of quantum theory, anti-correlation is observable from
classical chaotic light, if
\begin{align}\label{Quantum-1}
\Psi_{\alpha\beta} = \frac{1}{\sqrt{2}}
\big{[}\mathscr{A}(\mathbf{r}_{\alpha 1}, t_1; \mathbf{r}_{\beta 2},
t_2) - \mathscr{A}(\mathbf{r}_{\beta 1}, t_1; \mathbf{r}_{\alpha 2},
t_2) \big{]}
\end{align}
is achievable.  A number of experimental approaches may achieve the
two-photon destructive interference condition of
Eq.~(\ref{Quantum-1}). In this experiment, it is the beamsplitter
(BS2) that breaks the symmetry of the effective wavefunction in
Eq.~(\ref{symme}) and introduces the ``$-$" sign in achieving
Eq.~(\ref{Quantum-1}). In fact, both destructive (``$-$") and
constructive (``$+$") two-photon interferences have been achieved in
the measurement of entangled two-photon states \cite{HOMAS}.  The
anti-correlation ``dip" has been widely used for identifying
nonclassical states since the mid 1980's \cite{Loudon}.

\begin{figure}[hbt]
\centering
\includegraphics[width=70mm]{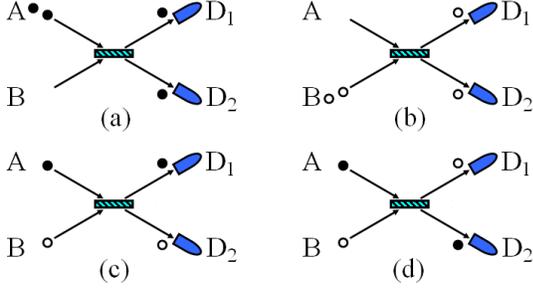}
\caption{There are four alternative ways for a measured pair of
independent photons to trigger a joint-detection event of $D_1$ and
$D_2$. } \label{fig:paths}
\end{figure}

In the view of quantum theory, either the correlation or the
anti-correlation of thermal radiation is the result of
\emph{two-photon interference}, which involves the nonlocal
superposition of two-photon amplitudes \cite{tpith}. Analogous to
Dirac's statement that a photon interferes with itself, this
interference is a jointly measured pair of independent photons
interfering with the pair itself. Figure~\ref{fig:paths}
schematically illustrates four alternatives for two independent
photons to trigger a joint-detection event of $D_1$ and $D_2$. In
(a) and (b) the measured pair comes from the same fiber tip, $A$ or
$B$. In (c) and (d) the measured pair comes from different fiber
tips, one from $A$ and the other from $B$.  It is the
superposition between amplitudes in (c) and (d) that produces the
anti-correlation.

Now we give a formal quantum mechanical calculation.  The probability
of observing a joint-detection event at space-time coordinates
$(\mathbf{r}_1, t_1)$ and $(\mathbf{r}_2,t_2)$ is given in
Eq.~(\ref{Quantum-0}), in which the fields $E(\mathbf{r}_1,t_1)$ and
$E(\mathbf{r}_2,t_2)$ are treated as the superposition of the early
fields at the input planes $z = d_A$ and $z = d_B$
\begin{align}\label{Field}
\hat{E}^{(+)}(z_1,t_1)&=\frac{1}{\sqrt{2}}[\hat{E}^{(+)}(\tau_{A1})
+ \hat{E}^{(+)}(\tau_{B1})] \nonumber \\
\hat{E}^{(+)}(z_2,t_2)&=\frac{1}{\sqrt{2}}[\hat{E}^{(+)}(\tau_{A2})
- \hat{E}^{(+)}(\tau_{B2})].
\end{align}
where $\tau_{Aj} \equiv t_j - r_{Aj}/c$, $j =1,2$,  the optical
delay from the detector $D_j$ to the input plane $A$, $n$ the index
of refraction of the fiber, and $\tau_{Bj}$ is defined similarly.

To simplify the calculation, we assume two groups of randomly
distributed wavepackets are excited at the fiber tips $A$ and $B$
\begin{align}
|\Psi_A \rangle &= \int d\omega \, f(\omega) \,
e^{i\omega t_{0A}} \, \hat{a}^{\dag}(\omega) \, |0\rangle
\nonumber \\
|\Psi_B \rangle &= \int d\omega \, f(\omega) \,
e^{i\omega t_{0B}} \, \hat{b}^{\dag}(\omega) \, |0\rangle,
\end{align}
where $f(\omega)$ is the spectrum function which is mainly
determined by the spectral filters (IF) in the experimental setup,
$t_{0A}$ and $t_{0B}$ represent the times when the wavepackets pass
the planes $A$ and $B$, respectively.

Substituting the field and the density operators into
Eq.~(\ref{Quantum-0}), it is straightforward to find that
\begin{align}\label{Quantum-10}
& \ \ \ \ G^{(2)}(z_1, t_1; z_2,t_2) \nonumber \\
&=|\mathscr{A}(\tau^T_{A1},\tau^R_{A2})|^2
+|\mathscr{A}(\tau^R_{B1},\tau^T_{B2})|^2 \nonumber \\
&\ \ \ +\left|\mathscr{A}(\tau^T_{A1},\tau^T_{B2})
-\mathscr{A}(\tau^R_{A2},\tau^R_{B1})\right|^2,
\end{align}
with
\begin{align*}
\mathscr{A}(\tau^T_{A1},\tau^R_{A2}) &= \langle 0 |
\hat{E}^{(+)}(\tau_{A1}) | \Psi_A \rangle  \langle 0 |
\hat{E}^{(+)}(\tau_{A2}) |\Psi_A \rangle \cr
\mathscr{A}(\tau^R_{B1},\tau^T_{B2}) &= \langle 0 |
\hat{E}^{(+)}(\tau_{B1}) | \Psi_B \rangle  \langle 0 |
\hat{E}^{(+)}(\tau_{B2}) |\Psi_B \rangle \cr
\mathscr{A}(\tau^T_{A1},\tau^T_{B2}) &= \langle 0 |
\hat{E}^{(+)}(\tau_{A1}) | \Psi_A \rangle  \langle 0 |
\hat{E}^{(+)}(\tau_{B2}) |\Psi_B \rangle \cr
\mathscr{A}(\tau^R_{B1},\tau^R_{A2}) &= \langle 0 |
\hat{E}^{(+)}(\tau_{B1}) | \Psi_B \rangle  \langle 0 |
\hat{E}^{(+)}(\tau_{A2}) |\Psi_A \rangle,
\end{align*}
where, $\tau^{T}_{A1} = (t_1- t_{0A}) - r_{A1}/c$ ($\tau^{R}_{B1} =
(t_1- t_{0B}) - r_{B1}/c$) indicating the early time at points $A$
($B$) by propagating through the reflected (transmitted) optical
path from source $A$ ($B$) to the photo-detector $D_1$;  and
$\tau^{R}_{A2} = (t_2- t_{0A}) - r_{A2}/c$ ($\tau^{T}_{B2} = (t_2-
t_{0B}) - r_{B2}/c$), indicating the early time at points $A$ ($B$)
by propagating through the transmitted (reflected) optical path from
$A$ ($B$) to the photo-detector $D_2$; $t_{0A}$ and $t_{0B}$ are the
initial times at $A$ and $B$, respectively. These four effective
wave functions correspond to the four alternatives shown in
Fig.\ref{fig:paths}.  It is not difficult to see that the first two
terms in Eq.~(\ref{Quantum-10}) are $\delta$ independent.  The
two-photon interference occurs in the third term.  The cross
two-photon interference term in the third term gives
$$
\mathscr{A}^*(\tau^T_{A1},\tau^T_{B2})\mathscr{A}(\tau^R_{B1},\tau^R_{A2})
\simeq e^{-\delta^{2}/\tau_c^{2}}
$$
for a Gaussian spectrum function of $f(\omega)$.  The details are given in Appendix B.
The coincidence coincident counting rate is therefore,
\begin{equation}\label{g22-2}
R_c \simeq R_0 \int dt_1 dt_2 \, G^{(2)}(z_1, t_1; z_2, t_2) \propto
1 - \frac{1}{2}e^{-\delta^{2}/\tau_c^{2}}.
\end{equation}
Eq.~(\ref{g22-2}) has been verified by achieving different coherence
time of the chaotic filed, which are determined by different
bandwidth of the spectral filters (IF), as shown in
Fig.\ref{fig:anti-correlation} (I) and (II), except with lower
contrasts [$\sim 28\%$ for (I) and $\sim 29\%$ for (II)].

In conclusion, we have observed a nonclassical anti-correlation from
classical chaotic light under the experimental condition of $G^{(1)}_{AB} =0$.
The classical statistical correlation theory seems unable to
explain this experimental result. This observation is different from all historical
measurement of the ``dips", either observed from nonclassical sources or from
synchronized coherent sources.  In the view of quantum mechanics,
either the anti-correlation ``dip" or the correlation ``peak" of
thermal light are straightforward two-photon interference phenomena,
involving the superposition of two-photon amplitudes.

The authors wish to thank T.B. Pittman, M.H. Rubin, J.P. Simon, Y.
Zhou, G. Scarcelli and V. Tamma for helpful discussions. This
research was partially supported by AFOSR and ARO-MURI program.  Z.
Xie acknowledges his partial support from China Scholarship Council.

\appendix
\section{Appendix A: Classical Statistical Intensity Correlation}
The standard statistics of classical chaotic light gives the second order
correlation in Eq.~(\ref{Classical-00}).  Eq.~(\ref{Classical-00}) can be
explicitly calculated as
\begin{align}\label{Classical-10A}
&\ \ \ \ \Gamma^{(2)}(\mathbf{r}_1, t_1; \mathbf{r}_2, t_2)
=\langle I(\mathbf{r}_1, t_1) \, I(\mathbf{r}_2, t_2) \rangle  \nonumber \\
&=\langle \big{|} E(\mathbf{r}_A, t_{1A}) + E(\mathbf{r}_B, t_{1B})
\big{|}^2
\big{|} E(\mathbf{r}_A, t_{2A}) - E(\mathbf{r}_B, t_{2B}) \big{|}^2 \rangle \nonumber \\
&=\langle \big{|} E(\mathbf{r}_A, t_{1A})\big{|}^{2} \rangle \, \langle \big{|} E(\mathbf{r}_A, t_{2A})\big{|}^{2} \rangle  \nonumber \\
&\ \ \ +\langle E^{*}(\mathbf{r}_A, t_{1A}) E(\mathbf{r}_A, t_{2A})
\rangle
\langle E(\mathbf{r}_A, t_{1A}) E^{*}(\mathbf{r}_A, t_{2A}) \rangle \nonumber \\
&\ \ \ +\langle \big{|} E(\mathbf{r}_B, t_{1B})\big{|}^{2} \rangle \, \langle \big{|} E(\mathbf{r}_B, t_{2B})\big{|}^{2} \rangle  \nonumber \\
&\ \ \ +\langle E^{*}(\mathbf{r}_B, t_{1B}) E(\mathbf{r}_B, t_{2B})
\rangle
\langle E(\mathbf{r}_B, t_{1B}) E^{*}(\mathbf{r}_B, t_{2B}) \rangle \nonumber \\
&\ \ \ +\langle \big{|} E(\mathbf{r}_A, t_{1A})\big{|}^{2} \rangle \, \langle \big{|} E(\mathbf{r}_B, t_{2B})\big{|}^{2} \rangle  \nonumber \\
&\ \ \ +\langle \big{|} E(\mathbf{r}_B, t_{1B})\big{|}^{2} \rangle \, \langle \big{|} E(\mathbf{r}_A, t_{2A})\big{|}^{2} \rangle  \\ \nonumber
&\ \ \ -\langle E^{*}(\mathbf{r}_A, t_{1A}) E(\mathbf{r}_A, t_{2A})
\rangle
\langle E(\mathbf{r}_B, t_{1B}) E^{*}(\mathbf{r}_B, t_{2B}) \rangle \\ \nonumber
&\ \ \ -\langle E(\mathbf{r}_A, t_{1A}) E^{*}(\mathbf{r}_A, t_{2A})
\rangle \langle E^{*}(\mathbf{r}_B, t_{1B}) E^(\mathbf{r}_B, t_{2B})
\rangle
\end{align}
where, again, $(\mathbf{r}_A, t_{jA} = t_j - r_{jA}/c)$ and
$(\mathbf{r}_B, t_{jB} = t_j - r_{jB}/c$), $j = 1,2$, are the
earlier space-time coordinates. Eq.~(\ref{Classical-10A}) can be
written in the following form
\begin{align}\label{Classical-11A}
&\Gamma^{(2)}=
\Gamma^{(1)}_{A11}\,\Gamma^{(1)}_{A22}+\Gamma^{(1)}_{A12}\,\Gamma^{(1)}_{A21}
+\Gamma^{(1)}_{B11}\,\Gamma^{(1)}_{B22}+\Gamma^{(1)}_{B12}\,\Gamma^{(1)}_{B21} \nonumber \\
&+\Gamma^{(1)}_{A11}\,\Gamma^{(1)}_{B22}+\Gamma^{(1)}_{A22}\,\Gamma^{(1)}_{B11}
-\Gamma^{(1)}_{A12}\,\Gamma^{(1)}_{B21}-\Gamma^{(1)}_{A21}\,\Gamma^{(1)}_{B12}.
\end{align}
Here, all the terms $\Gamma^{(1)}_{AB}\Gamma^{(1)}_{AB}$ vanish due
to mutual incoherence nature between the fields $E(\mathbf{r}_A,
t_A)$ and  $E(\mathbf{r}_B, t_B)$, leaving the non-zero terms
$\Gamma^{(1)}_{AA}\Gamma^{(1)}_{AA}$,
$\Gamma^{(1)}_{BB}\,\Gamma^{(1)}_{BB}$ and
$\Gamma^{(1)}_{AA}\Gamma^{(1)}_{BB}$ contributing to
$\Gamma^{(2)}$. The negative signs in the last two terms are
basically introduced by beamsplitter. It is not too difficult to find from Eq.~(\ref{Classical-10A})
that $\Gamma^{(2)}$ is independent of the optical delay $\delta$ for either CW or pulsed
chaotic light.

\vspace{3mm}
\hspace{-4.5mm}Case (I): CW chaotic light
\vspace{3mm}

There is no doubt the first four terms are $\delta$ independent.  In
these terms either the self-correlations, $ \Gamma^{(1)}_{A11}
\Gamma^{(1)}_{A22}$ and $\Gamma^{(1)}_{B11} \Gamma^{(1)}_{B22}$, or
the cross-correlations, $\Gamma^{(1)}_{A12} \Gamma^{(1)}_{A21}$ and
$\Gamma^{(1)}_{B12} \Gamma^{(1)}_{B21}$ are respectively associated
with the same radiation $A$ or $B$.   The other four terms may contain
$\delta$ in their amplitude part or in their phase part.  We do not
need to worry about the amplitude part due to the stationary nature
of the CW chaotic light.   Let us examine the phase part:  (1) the
first two terms $\Gamma^{(1)}_{A11} \Gamma^{(1)}_{B22} = \langle
I_{A11} \rangle \langle I_{B22} \rangle$ and $\Gamma^{(1)}_{A22}
\Gamma^{(1)}_{B11} = \langle I_{A22} \rangle \langle I_{B11}
\rangle$ contain the expectation of intensities only and thus are
phase independent. (2)The second two terms $\Gamma^{(1)}_{A12}
\Gamma^{(1)}_{B21}$ and $\Gamma^{(1)}_{A21}  \Gamma^{(1)}_{B12}$ may
contain relative phases of $k(r_{A1} - r_{A2})$ and $k(r_{B1} -
r_{B2})$, however, these phases are $\delta$ independent.   We may
conclude in the case of CW chaotic light,  $\Gamma^{(2)}$ is
independent of the optical delay $\delta$.

\vspace{3mm}
\hspace{-4.5mm}Case (II): Pulsed chaotic light
\vspace{3mm}

All the analysis are the same as above, except we do need to consider the delays between
the amplitudes, which is $\delta$ dependent.  In the pulsed case, the fields are non-stationary,
the last four terms have nonzero values only when their amplitudes have nonzero
values simultaneously in the joint-detection of $D_1$ and $D_2$.  It is clear when $\delta =0$,
all four terms achieve their maximum values simultaneously due to the complete overlapping of
their respective amplitudes.   The overall contribution of the four terms, however,
is null to $\Gamma^{(2)}$ due to the cancelation between the
first two positive contributions and the last two negative contributions.  When $\delta$ increase
or decrease from zero, although the overlapping between the $A$ amplitude and the $B$ amplitude
becomes smaller and smaller yielding smaller contributions to $\Gamma^{(2)}$, the overall
contribution keeps its zero vale for the same reason.  We may
conclude that in the case of pulsed chaotic light, $\Gamma^{(2)}$ is independent
of the optical delay $\delta$ too.

\section{Appendix B: Quantum Theory}
In Appendix B we show a simple calculation of the effective wave functions.
The effective-wavefunctions corresponding to the case in which the joint-detection
event is produced by two photon coming from the same point is shown below as
an example
\begin{align}
& \ \ \ \mathscr{A}(\tau^T_{A1},\tau^R_{A2}) = \langle 0 |
\hat{E}^{(+)}(\tau_{A1}) | \Psi_A \rangle  \langle 0 |
\hat{E}^{(+)}(\tau_{A2}) |\Psi_A \rangle \nonumber \\
&\propto\int d\omega f(\omega) e^{-i\omega\tau^T_{A1}}\int d\omega'
f(\omega')
e^{-i\omega'\tau^R_{A2}}\nonumber \\
&=e^{-i\omega_0
\tau^T_{A1}}\mathcal{F}_{\tau^T_{A1}}\{f(\nu)\}e^{-i\omega_0
\tau^R_{A2}}\mathcal{F}_{\tau^R_{A2}}\{f(\nu)\}
\end{align}
where,
$$
\mathcal{F}_\tau\{f(\nu)\}\equiv\int d\nu f(\nu) e^{-i\nu \tau}
$$
is the fourier transform of the spectrum function $f(\nu)$,
and $\nu=\omega-\omega_0$ ($\omega_0$ is the central frequency).
%\begin{align*}
%&\int_T dt_1 dt_2
%\left|\mathscr{A}(\tau^T_{A1},\tau^R_{A2})\right|^2 \cr &\propto\int
%dt_1 dt_2
%\left|\mathcal{F}_{\tau^T_{A1}}\{f(\nu)\}\mathcal{F}_{\tau^R_{A2}}\{f(\nu)\}\right|^2\cr
%&=I_0^2
%\end{align*}
%where $I_0\equiv\int d\nu f(\nu)$.
It is easy to see these functions are $\delta$ independent.

The superposition of the two-photon amplitudes (c) and (d) corresponding to
the interference term in Eq.~(\ref{Quantum-10}) is calculated to be
\begin{align}\label{Quantum-10A}
& \ \ \ \ \big{|} \mathscr{A}(\tau^T_{A1},
\tau^T_{B2}) - \mathscr{A}(\tau^R_{B1}, \tau^R_{A2}) \big{|}^2 \\ \nonumber
%&=\left|\mathcal{F}_{\tau^T_{A1}}\big{\{} f(\nu)
%\big{\}}\mathcal{F}_{\tau^T_{B2}}\big{\{} f(\nu) \big{\}}\right.\cr&
% \left.-\mathcal{F}_{\tau^R_{B1}}\big{\{}
%f(\nu) \big{\}}\mathcal{F}_{\tau^R_{A2}}\big{\{} f(\nu)
%\big{\}}\right|^2 \cr
&= \big{|}\mathcal{F}_{\tau^T_{A1}}\big{\{} f(\nu) \big{\}}
\mathcal{F}_{\tau^T_{B2}}\big{\{} f(\nu) \big{\}}  \big{|}^2 +
\big{|} \mathcal{F}_{\tau^R_{B1}}\big{\{} f(\nu) \big{\}}
\mathcal{F}_{\tau^R_{A2}}\big{\{} f(\nu) \big{\}}  \big{|}^2 \\
\nonumber & \ \ \ - \mathcal{F}^*_{\tau^T_{A1}}\big{\{} f(\nu)
\big{\}} \mathcal{F}^*_{\tau^T_{B2}}\big{\{} f(\nu) \big{\}}
\mathcal{F}_{\tau^R_{B1}}\big{\{} f(\nu) \big{\}}
\mathcal{F}_{\tau^R_{A2}}\big{\{} f(\nu) \big{\}} \\ \nonumber & \ \
\ -  \mathcal{F}_{\tau^R_{A1}}\big{\{} f(\nu) \big{\}}
\mathcal{F}_{\tau^T_{B2}}\big{\{} f(\nu) \big{\}}
\mathcal{F}^*_{\tau^R_{B1}}\big{\{} f(\nu) \big{\}}
\mathcal{F}^*_{\tau^R_{A2}}\big{\{} f(\nu) \big{\}}.
\end{align}
It is easy to find that the first two terms in Eq.~(\ref{Quantum-10A}) contribute
a constant to the coincidence counting rate $R_c$. The nontrivial contributions
come from the last two cross terms.  Assuming a Gaussian spectrum, the cross terms
is approximately to be
\begin{equation*}
\mathcal{F}^*_{\tau^T_{A1}}\big{\{} f(\nu) \big{\}}
\mathcal{F}^*_{\tau^T_{B2}}\big{\{} f(\nu) \big{\}}
\mathcal{F}_{\tau^R_{B1}}\big{\{} f(\nu) \big{\}}
\mathcal{F}_{\tau^R_{A2}}\big{\{} f(\nu) \big{\}} \propto  e^{-
\delta^2 / \tau^2_c},
\end{equation*}
for synchronized radiations $A$ and $B$, i.e., $t_{0A} = t_{0B}$, where
$\tau_c$ is the coherence time of the measured field.  It is
interesting to see the cross term contribution is time independent.
The coincidence counting rate is therefore
\begin{equation*}
R_c \simeq R_0 \int dt_1 dt_2 \, G^{(2)}(z_1, t_1; z_2, t_2) \propto
1 - \frac{1}{2}e^{-\delta^{2}/\tau_c^{2}}.
\end{equation*}


\begin{thebibliography}{}

\bibitem{HBT} Hanbury Brown R, and Twiss R Q 1956 {\it
Nature} {\textbf 177}, 27-29; Hanbury Brown R 1974 \emph{Intensity
Interferometer} (Taylor \& Francis, London).

\bibitem{Gpaper}
Scarcelli G, Berardi V, and Shih Y H 2006 {\it Phys. Rev. Lett.} \textbf{96},
063602.  Note, the natural, non-factorizable, point-to-point, near-field-image-forming
correlation in the lensless ghost imaging is in principle different from these
classical simulations, such as the man-made, factorizable, ``speckle to speckle",
correlation of A. Gatti \emph{et al}. [A.~Gatti \emph{et al}, Phys. Rev. A \textbf{70},
013802, (2004), and Phys. Rev. Lett. \textbf{93}, 093602 (2004)]. The original
publications of Gatti \emph{et al}. choose 2f-2f classical imaging systems,
$1/2f + 1/2f = 1/f$, to image the ``speckles" of the light source onto the object
plane and the ghost image plane, respectively.  Their speckle-speckle correlation
is factorizeable into two classical images of ``speckles".


\bibitem{Mandel} $\Gamma^{(1)}_{AB} = 0$ makes
this experiment different from all other demonstrations of
interference between independent light sources. Mandel \emph{et al.}
showed that radiations from two independent sources can produce
first-order interference for an single-shot exposure, if the
``exposure" time is shorter than the coherence time of the fields.
See, Magyar G, Mandel L 1963 {\it Nature}, \textbf{198}, 255; Mandel
L 1983 {\it Phys. Rev. A}, \textbf{28}, 929. In a time integrated
multi-exposure measurement, the first-order interference pattern may
not be observable due to the phase variation from one exposure to
another exposure.  However, the joint-detection of two individual
photodetectors may still produce observable second-order correlation
or anti-correlation.  In our experiment, $\Gamma^{(1)}_{AB}$ is
chosen to be zero.  The classical interpretation of time averaged
effect of classical mutual coherence or partial mutual coherence
does not apply.

\bibitem{Pseudo} Martienssen W and Spiller E 1964 {\it Am. J. Phys.}
\textbf{32}, 919.

\bibitem{Glauber} Glauber R J 1963 {\it Phys. Rev.} \textbf{10} 84;
Glauber R J 1963 {\it Phys. Rev.} \textbf{130}, 2529.  In
Eq.~(\ref{Quantum-0}), we use $G^{(2)}$ to distinguish quantum
correlation from classical statistical intensity correlation of
$\Gamma^{(2)}$.

\bibitem{Bari} Jeltes T, et al. 2007
{\it Nature} 445, 402-405.

\bibitem{HOMAS} Hong C K, Ou Z Y  and Mandel L 1987 {\it Phys. Rev. Lett.} \textit{59}, 2044;
Shih Y H and Alley C O 1988 {\it Phys. Rev. Lett.} \textit{61},
2921.  In the Alley-Shih two-photon interferometer, both
anti-correlation ``dip" and correlation ``peak" are observable by
selecting different polarization of the entangled photon pair.

\bibitem{Loudon} Loudon R 2000 \emph{The Quantum Theory of
Light} (Oxford, New York, 3rd Edition ).

\bibitem{tpith}G. Scarcelli, A. Valencia
and Shih Y H 2004 {\it EuroPhys. Lett.} {\textbf 68}: 618.

\end{thebibliography}
\end{document}